\documentclass[sigplan, twocolumn]{acmart}
\acmSubmissionID{58}
\renewcommand\footnotetextcopyrightpermission[1]{}
\renewcommand\footnotetextcopyrightpermission[1]{}
\acmConference[]{}{}{}

\usepackage{tikz}
\usepackage{amsmath}
\usepackage{graphicx}
\usepackage{subfigure}
\usepackage{amsmath}
\usepackage{soul}
\usepackage{pifont}
\usepackage{enumitem}

\newcommand{\note}[1]{{{\textcolor{blue}{#1}}}}
\newcommand{\name}[1]{\emph{vLSM}}
\newcommand{\nameb}[1]{\emph{TriLSM}}
\newcommand{\betree}[1]{{B$^{\epsilon}$--Tree}}
\newcommand{\commentout}[1]{{}}
\newif\ifshowcomment
\showcommenttrue

\newcommand{\mathvar}[1]{\text{\emph{#1}}}
\ifshowcomment
\newcommand{\todo}[1]{\noindent\textsf{\color{orange}{[{todo: \it #1}]}}}
\newcommand{\ant}[1]{\noindent\textsf{\color{purple}{[ak: {\it#1}]}}}
\newcommand{\antonis}[1]{\noindent\textsf{\color{purple}{[ak: {\it#1}]}}}
\else
\newcommand{\ant}[1]{}
\newcommand{\antonis}[1]{}
\newcommand{\todo}[1]{}
\fi
\newcommand*\circled[1]{\tikz[baseline=(char.base)]{
   \node[shape=circle,draw,fill=gray!40,inner sep=0.5pt](char){\textcolor{black}{#1}};}}

\newcommand*\scircled[1]{
\scalebox{0.8}{
\tikz[baseline=(char.base)]{
            \node[shape=circle, text=white, fill=black, draw,inner sep=0.5pt] (char) {#1};}}}

\makeatletter
\def\hlinewd#1{%
\noalign{\ifnum0=`}\fi\hrule \@height #1 %
\futurelet\reserved@a\@xhline}
\makeatother

\def\custspace{\vspace{3pt}}
\newcommand{\beginbsec}[1]{\custspace\noindent\textbf{#1. \hspace{2pt}}}

\usepackage[capitalize, noabbrev, nameinlink]{cleveref}
\crefname{section}{\S}{\S}
\Crefname{section}{\S}{\S}
\author{Giorgos Xanthakis}
\email{gxanth@ics.forth.gr}
\affiliation{%
   \institution{University of Crete \& ICS-FORTH}
   \city{Heraklion}
   \country{Greece}
}

\author{Antonios Katsarakis$^1$}
\email{antonios.katsarakis@huawei.com}
\affiliation{%
   \institution{Huawei}
   \city{Edinburgh}
   \country{United Kingdom}
}

\author{Giorgos Saloustros}
\email{gesalous@ics.forth.gr}
\affiliation{%
   \institution{University of Crete \& ICS-FORTH}
   \city{Heraklion}
   \country{Greece}
}

\author{Angelos Bilas}
\email{bilas@ics.forth.gr}
\affiliation{%
   \institution{University of Crete \& ICS-FORTH}
   \city{Heraklion}
   \country{Greece}
}

\begin{document}
\date{}
\title[vLSM: Low tail latency and I/O amplification in LSM-based KV stores]{
   \scalebox{0.78}{
      vLSM: Low tail latency and I/O amplification in LSM-based KV stores}}

\begin{abstract}

    LSM-based key-value (KV) stores are an important component in modern data infrastructures. However, they suffer from high tail
    latency, in the order of several seconds, making them less attractive for user-facing applications.

    In this paper, we introduce the notion of compaction chains and we analyse how they affect tail latency. Then, we show that modern
    designs reduce tail latency, by trading I/O amplification or require large amounts of memory.

    Based on our analysis, we
    present \name{}, a new KV store design that improves tail latency significantly without compromising on memory or I/O
    amplification. \name{} reduces (a) compaction chain width by using small SSTs and eliminating the tiering compaction required in
    $L_0$ by modern systems and (b) compaction chain length by using a larger than typical growth factor between $L_1$ and $L_2$ and
    introducing overlap-aware SSTs in $L_1$.

    We implement \name{} in RocksDB and evaluate it using db\_bench and YCSB. Our evaluation highlights the underlying trade-off among
    memory requirements, I/O amplification, and tail latency, as well as the advantage of \name{} over current approaches. \name{}
    improves P99 tail latency by up to 4.8$\times{}$ for writes and by up to 12.5$\times{}$ for reads, reduces cumulative write stalls
    by up to 60\% while also slightly improves I/O amplification at the same memory budget.
    \footnotetext[1]{This work was done while the author was at the University of Edinburgh.}
\end{abstract}

\maketitle
\sloppy
\section{Introduction}
\label{sec:intro}

Log-structured merge-tree (LSM) key-value (KV) stores are a cornerstone in the evolution of modern storage
systems~\cite{rocksdbtos,bigtable,cassandra}.  They are the backbone of popular user-facing applications and services, including
social media~\cite{fbookstudy, discord}, financial systems~\cite{tigerbeetle}, and AI workflows~\cite{aiworkflows}. Such
applications demand cost-effective access to large volumes of data via highly concurrent and latency-sensitive requests.

To deliver high throughput cost-effectively, state-of-the-art LSM KV stores optimize for low I/O amplification and low
memory usage. To achieve this, modern KV stores introduce a tiering step in the first level on the device
($L_0$)~\cite{rocksdbtiering}. This tiering step allows them to use a small amount of memory for the in-memory component, while
$L_0$ can still be large. This results in an LSM with fewer levels and, therefore, reduced I/O amplification. However, the tiering
compaction step leads to prolonged write-stalls, which inflate tail latency~\cite{matrixkv2, novelsm}.

\begin{figure}[t]
  \subfigure[Write stalls]{\includegraphics[width=.49\linewidth]{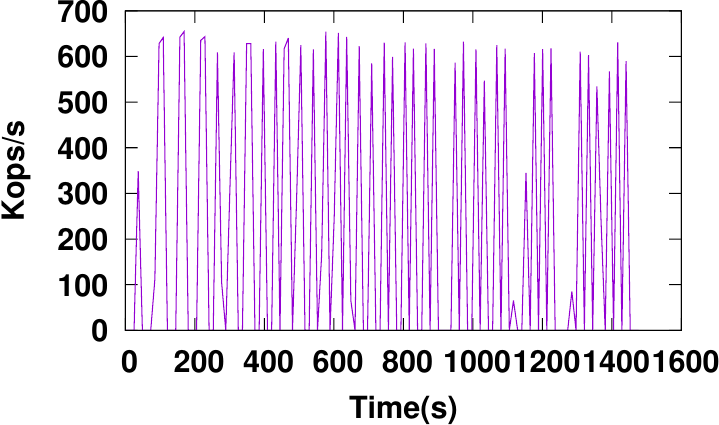}\label{fig:stalls}}
  \hfill
  \subfigure[P99 latency]{\includegraphics[width=0.47\linewidth]{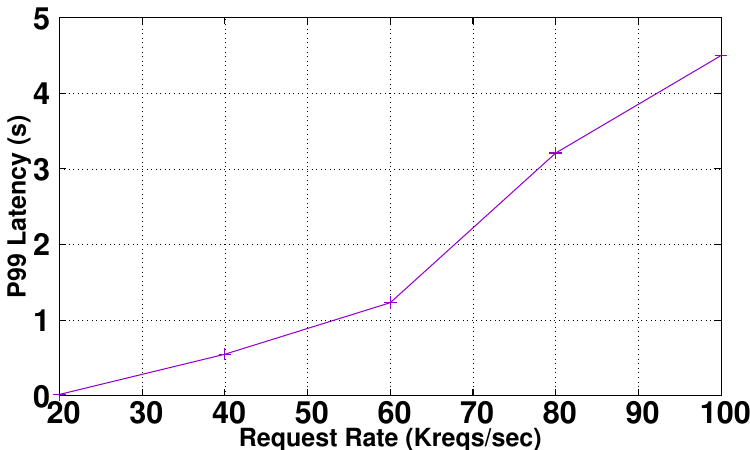}\label{fig:reqratelatintro}}
  \caption{RocksDB throughput and P99 latency in YCSB Load A, using 350M KV pairs with a KV size of 240 B.}
\end{figure}

Figure~\ref{fig:stalls} shows RockDB's throughput over time for YCSB (Load A -- more details in~\cref{sec:method}). The runtime is
significantly impacted by write stalls, which occur due to compaction chains that must occur to free space before RocksDB can
ingest new requests.  These write stalls account for approximately 40\% of the total runtime and drastically affect tail latency
and the user experience. Figure ~\ref{fig:reqratelatintro} shows the 99th-percentile (P99) latency as the load increases in the
same setup. RocksDB's P99 latency is in the order of seconds, even with a load less than 60\% of its maximum throughput.

Related works have tried to mitigate the high tail latency caused by compactions using two different approaches:
\textit{memory-based} or \textit{scheduling-based} solutions. Memory-based solutions, keep significantly more data in memory,
i.e. mandate a high memory budget, to reduce the number of slower disk-resident compactions in the request critical path. These
solutions are costly and mainly rely on new memory technologies, such as byte-addressable non-volatile memory~\cite{matrixkv2,
  novelsm} that may not be broadly available~\cite{disnvm}.

In contrast, scheduling-based solutions aim to schedule compactions in the background to reduce the amount of work that needs to
occur in the critical path, and therefore, reduce tail latency~\cite{silk, blsm, adoc}. While these approaches can significantly
lower write stalls and tail latency, they assume a light load or over-provisioned resources (e.g., underutilized CPU or device
resources) to perform compactions in parallel with the running workload. More importantly, scheduling-based approaches increase I/O
amplification by temporarily increasing the size of levels, resulting in more background work per compaction. For example, our
evaluation shows that a state-of-the-art scheduling approach ADOC~\cite{adoc} increases I/O amplification from 26$\times$ to
46$\times$.

We identify two main factors in traditional LSM compactions that affect tail latency. Namely, the maximum \textit{width} and the
maximum \textit{length} of compaction chains that can occur on the critical path of requests. Simply put, the width is determined
by the amount of compacted bytes per level, while the length refers to the total number of levels that must be compacted to free memory. For
state-of-the-art LSM KV stores, the tiering compaction step in $L_0$ governs the width, which typically is 2 GBs per level
in RocksDB. Assuming we have 5 to 7 levels in a typical LSM KV store, the combined width and length of compaction chains reaches tenths of GBs. Intuitively, an LSM design should reduce those without inflating I/O amplification or memory usage.

Based on this insight, we introduce \name{}, a novel LSM design that ensures low tail latency and low I/O amplification at a small
memory budget. To achieve this, \name{} carefully reduces both the width and height of compaction chains in LSMs.
It reduces the width of compaction chains by\scircled{1} employing smaller SSTs and\scircled{2} removing tiering compactions. To
keep compaction chains short without compromising on memory or I/O amplification, \name{} meticulously handles the first
device-only levels ($L_1$ and $L_2$). It applies\scircled{3} a larger growth factor $\Phi$ from $L_1$ to $L_2$ and\scircled{4} introduces
overlap-aware variable-size SSTs (vSSTs) in $L_1$. The size of vSSTs in $L_1$, is determined by a novel look-ahead compaction
policy that minimizes their overlap with SSTs in $L_2$.

We evaluate \name{} against RocksDB and ADOC over standard workloads and benchmarks (YCSB and db\_bench) as
well as sensitivity studies.  Our experiments demonstrate substantial improvements in tail latency without compromising on I/O
amplification or memory size. Compared to RocksDB (ADOC), \name{} reduces P99 latency up to 5.2$\times$ (1.99$\times$) for YCSB
Load A that performs inserts. For mixed read-write workloads, such as YCSB Run A, \name{} improves the P99 write latency by
4.8$\times$ (2.13$\times$) and P99 read latency by 12.5$\times$. At the same time, \name{} requires a similar amount of memory as
existing designs that optimize for I/O amplification and memory use. Overall, our contributions are:

\begin{itemize}[leftmargin=*]
  \item \textbf{We identify that state-of-the-art memory-frugal LSM stores sacrifice tail latency or I/O amplification}. We
        pinpoint the root cause of this conundrum to the width and length of compaction chains that occur on the critical path of a
        request.
  \item \textbf{We introduce \name{}, a novel KV store design} that achieves low tail latency without inflating I/O amplification
        or memory usage. \name{} achieves this by reducing the width and length of compaction chains through the combination of four
        key ideas:\scircled{1} smaller SSTs,\scircled{2} eschewing tiering compactions,\scircled{3} larger growth factor from $L_1$ to $L_2$,
        and\scircled{4} overlap-aware vSSTs in $L_1$.
  \item \textbf{We extensively evaluate \name{}} using standard benchmarks and sensitivity studies. Our results show that \name{}
        shrinks compaction chains by up to 10$\times$, which translates into 4.8$\times$ (12.5$\times$) lower latency for writes
        (reads) over RocksDB and up to 1.7$\times$ lower I/O amplification than the state-of-the-art scheduling solution of ADOC.
\end{itemize}



\section{Motivation}
\label{sec:motivation}
\subsection{Modern KV Stores}
A KV store typically divides its data in \textit{regions}, often in the order of hundreds~\cite{rocksdbevol}. Each
region is a subset of a KV range with an independent LSM index from other regions. It has a predefined number of maximum levels
$n$, which is derived from its maximum capacity, the size of $L_0$, and the growth factor $f$ across levels. In each region, when
$L_0$ is full, the LSM KV store selects an SST from $L_0$ and compacts it with the overlapping SSTs from
$L_1$~\cite{incremental_compaction}.

\subsection{Incremental compactions}
Modern KV stores use incremental compaction~\cite{rocksdbtos, incremental_compaction}. They organize levels in non-overlapping
sub-units that contain sorted KV pairs named Sorted String Tables (SSTs). Assuming a constant growth factor $f$, KV stores organize
each level with fixed-size SSTs (e.g., 64 MB) and increase their number from level to level by $f$ times. It is important to note
that each SST is a self-contained unit, stored in a single file or extent on the device, containing all its data (KV pairs) and
metadata (index and bloom filters).

SSTs enable incremental compaction. They allow KV stores to merge-sort part of a level (one or a few SSTs) into
the next level.
Simply, when a level is full, instead of compacting the whole level, the KV store can compact a single SST to the next
level. Therefore, the LSM KV store must reorganize $n$ levels touching only $n\times f$ SSTs to free up space for a subsequent insert operation,
dramatically reducing the amount of work and decreasing tail latency compared to full compaction~\cite{lsm}.  Under incremental compaction,
the width of the compaction chains is affected by the growth factor $f$ and the size of SSTs. The growth factor indicates how many
SSTs of the target level are involved, on average, in an incremental compaction step. The size of SSTs affects the amount of work
required to merge each overlapping SST.


\subsection{Compaction chains}
In this work, we identify that the root cause of tail latency in modern LSM-based KV stores are \emph{compaction chains} that form during operation.
%
Compaction chains are sequences of dependent compactions from level to level. These compactions need to be processed in order to
maintain the correctness properties of the LSM and keep the latest data higher up in the tree.
Hence, when levels are full, we first need to free space in $L_{n-1}$, then $L_{n-2}$, and so on, until the in-memory component.

Compaction chains have two aspects that affect tail latency. First, the \emph{length} of the chain, which is defined by the number
of LSM levels in the chain. The number of LSM levels is defined by the size of the in-memory component and the growth factor $f$.
Second, the amount of work for each compaction in a chain, which we call the \emph{width} of the chain.
Figure~\ref{fig:design-vanilla}, illustrates the width and the length of compaction chains over an LSM with incremental compactions (LSMi).

A chain of dependent compactions from $L_0$ to $L_n$ must take place to free space in each next level whenever they are full. The
aggregate capacity of $L_0$ to $L_{n-1}$ level is about 10\% of the total KV store space for typical growth factors around
8-10~\cite{vat, spacerocks}. As a result,
long and fat compaction chains are observed as early as the dataset size reaches or
exceeds about 10\% of the capacity of each region.

\begin{figure}
  \subfigure[Chain width] {\includegraphics[width=0.49\columnwidth]{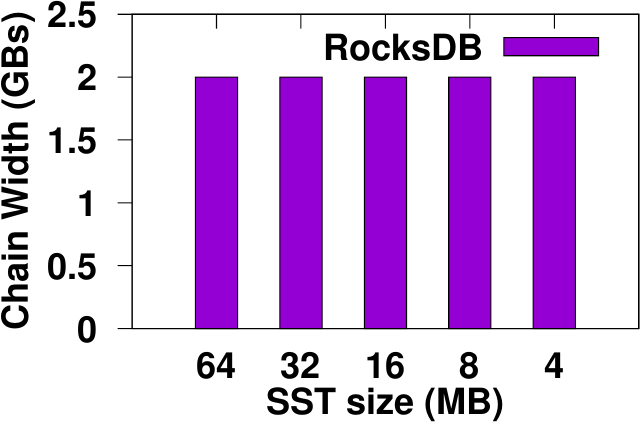}\label{fig:rdbcc-width}}
  \subfigure[Chain length] {\includegraphics[width=0.50\columnwidth]{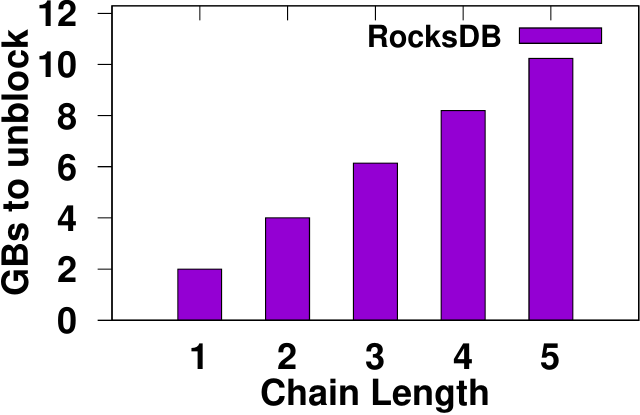}\label{fig:rdbcc-length}}
  \caption{RocksDB chain width (left) and chain length (right) for different SST sizes.}
\end{figure}

Figure~\ref{fig:rdbcc-width} shows how much data compaction moves
on average on each level. Figure~\ref{fig:rdbcc-length} shows the amount of data that needs to be compacted to free space for a memtable in
$L_0$ with a different number of levels in RocksDB. The width of the compaction chain is 2 GBs per level and increases by $L_0$ size
for each level. So, for a typical deployment with five levels, an LSM KV store needs to reorganize up to 10 GBs of data to free space for 64 MBs in memory, which is about 20 GBs of read/write traffic.

\section{Analysis of compaction chains:
  \scalebox{0.8}{Tail latency \textit{~vs.~} I/O amplification \textit{~vs.~} memory usage}
 }
\label{sec:chains}

We observe that state-of-the-art systems follow three strategies to reduce the cost of compaction chains and, thus, tail latency.

\begin{enumerate}[leftmargin=*]
  \item Reduce the length of chains by using a larger in-memory component at the cost of increased memory.
  \item Reduce the compaction width of each stage by using smaller SSTs, which leads to more levels.
  \item Use scheduling techniques to eliminate dependencies and execute compactions within a chain in parallel. The disadvantage of this approach is the violation of the growth factor, which increases I/O amplification.
\end{enumerate}

Next, we examine each strategy for improving tail latency, concluding that in all cases, there is a negative impact on the amount of
required memory or I/O amplification.

\subsection{Reducing the \textit{length} of compaction chains}
It is relatively straightforward to reduce the length of compaction chains by either increasing the growth factor $f$ or the size of $L_0$. However,
both of these result in a significant increase in I/O amplification or memory usage. Increasing the growth factor from 10 to 20
would reduce the number of levels by one at the cost of doubling the I/O amplification~\cite{vat}.

Alternatively, increasing $L_0$ by a growth factor $f$ (typically $8-10\times$) would reduce the number of levels by one and increase the compaction width $f$ times.
However, on servers that host several KV regions (typically hundreds or thousands~\cite{rocksdbevol}), this results in excessive memory usage up to $f$ times.

Another way to reduce the length of compaction chains is to break a dataset into a larger number of smaller regions. Small regions employ fewer levels for the same growth factor $f$. However, this approach similarly increases the amount of memory required, as
the collective in-memory component of all regions increases in size compared to using fewer regions. Therefore, such approaches cannot improve tail latency in a cost-effective manner.

A more practical approach to reduce the number of levels uses a tiering compaction step in $L_0$, as follows. The original
LSM-tree~\cite{lsm} design requires that the entire $L_0$ is in-memory. Modern KV stores, such as RocksDB~\cite{rocksdb}, deviate
from the original design to increase the size of $L_0$, which results in fewer levels and less I/O amplification.  However, to
avoid increasing memory requirements, they keep two versions of $L_0$: a tiered $L_0$ ($L_0$ in RocksDB terminology) and a leveled $L_0$
($L_1$ in RocksDB terminology), which are the same size.
Tiered $L_0$ consists of overlapping SSTs, whereas $L_1$ consists of
non-overlapping SSTs. Both $L_0$ and $L_1$ levels reside on the device without requiring memory. Then, it uses \emph{memtables} as
the in-memory component of the KV store. Each memtable, similar to SSTs, has a size of several tens of MB (64 by default).

Figure~\ref{fig:design-rdb}
shows the data path in RocksDB with incremental compactions and a tiering compaction step in $L_0$. When a memtable fills, the system converts
the memtable to SST format and flushes it to $L_0$. When $L_0$ is full, RocksDB moves one SST at a time using a
bottom-up approach from $L_{n-1}$ to $L_1$ to free enough space for the SSTs of $L_1$. Then RocksDB reads the SSTs of
$L_0$ and reorganizes them in non-overlapping units in-memory (tiering step). Finally, it flushes the resulting SSTs to the empty
$L_1$. With this tiering step, RocksDB can use a large $L_0$, in the order of GB, with just a few MB of memory budget
for memtables, leading to almost one order of magnitude memory savings compared to designs that use an in-memory $L_0$ (e.g. the
original LSM).

\subsection{Reducing the \textit{width} of compaction chains}
Reducing the width of compaction chains is more intricate. Traditionally, LSM KV stores use relatively large SSTs (e.g., 64 MBs) to cater to HDD
characteristics and reduce constant overheads in the KV store design, such as for managing guards in memory or for recovering from
failures. However, large SSTs significantly increase the width of compaction chains, resulting in reorganizing hundreds of MB in
each compaction step. Although the emergence of flash devices creates an opportunity to reduce SST size and compaction width, it
turns out
reducing the size of SSTs alone is not enough to
significantly improve tail latency.

\begin{figure*}[t]
  \centering
  \subfigure[LSMi]{\includegraphics[width=.245\linewidth]{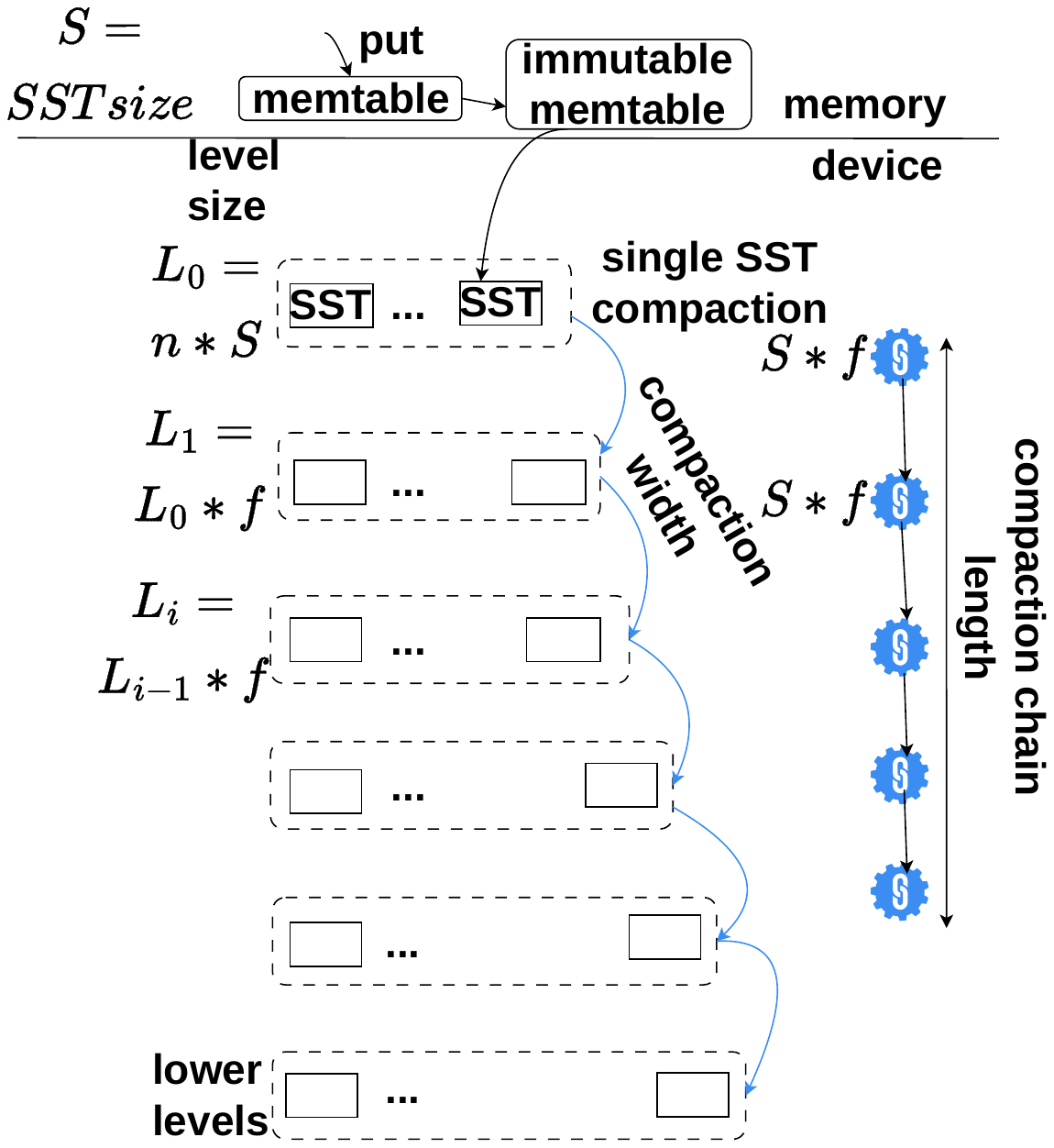}\label{fig:design-vanilla}}
  \subfigure[RocksDB-IO]{\includegraphics[width=.245\linewidth]{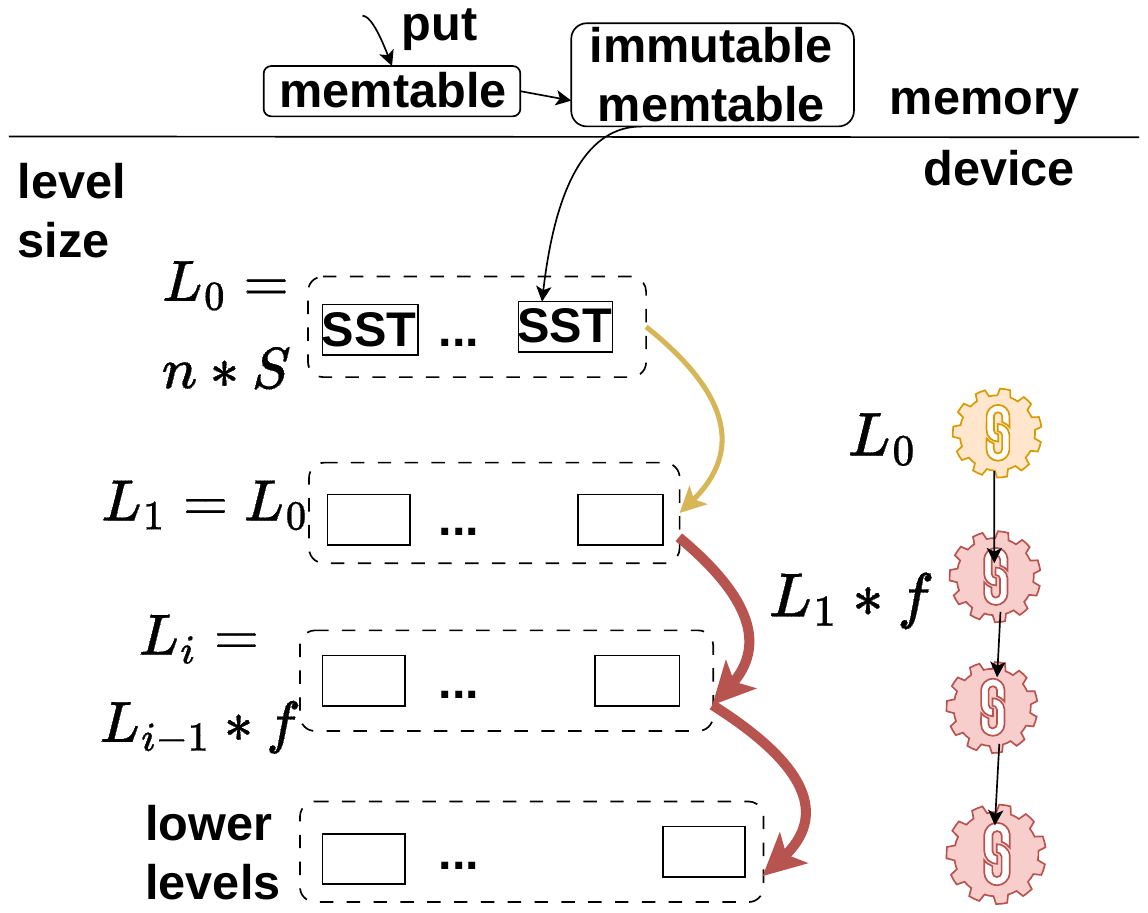}\label{fig:design-rdb}}
  \subfigure[ADOC (over RocksDB)]{\includegraphics[width=.245\linewidth]{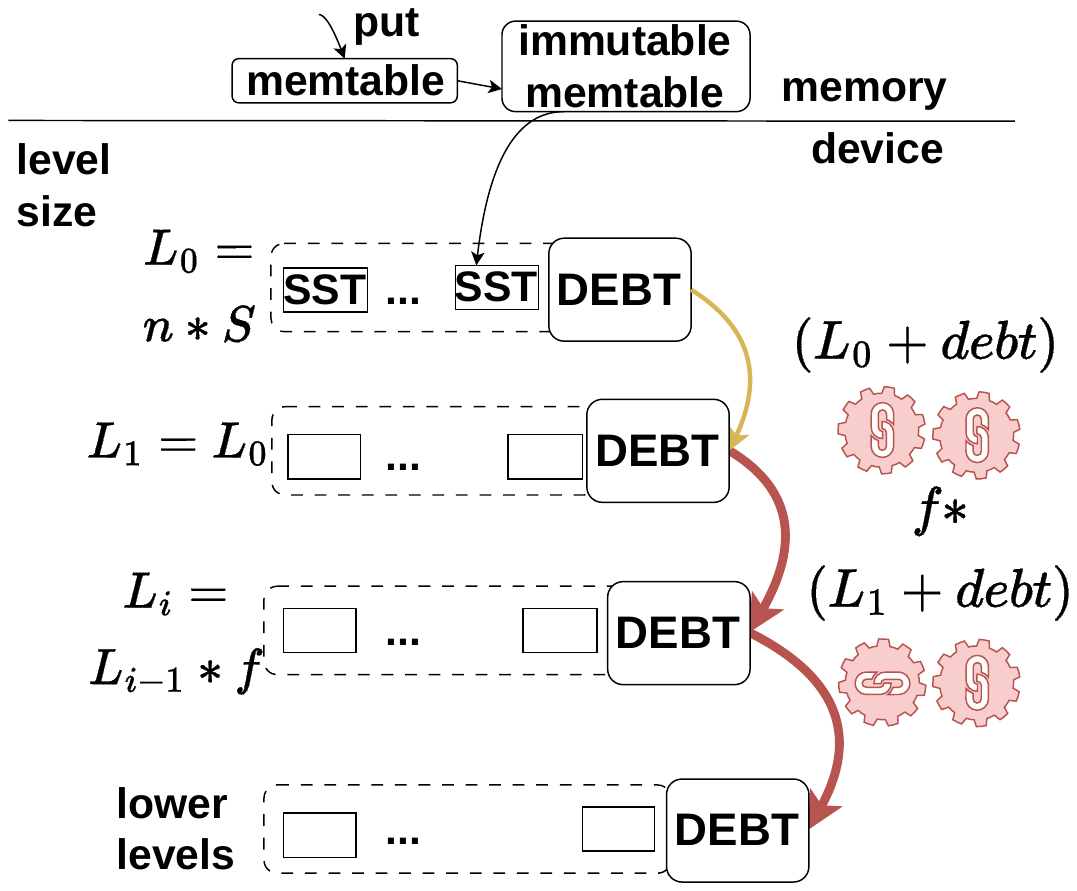}\label{fig:design-adoc}}
  \subfigure[\name{}]{\includegraphics[width=.245\linewidth]{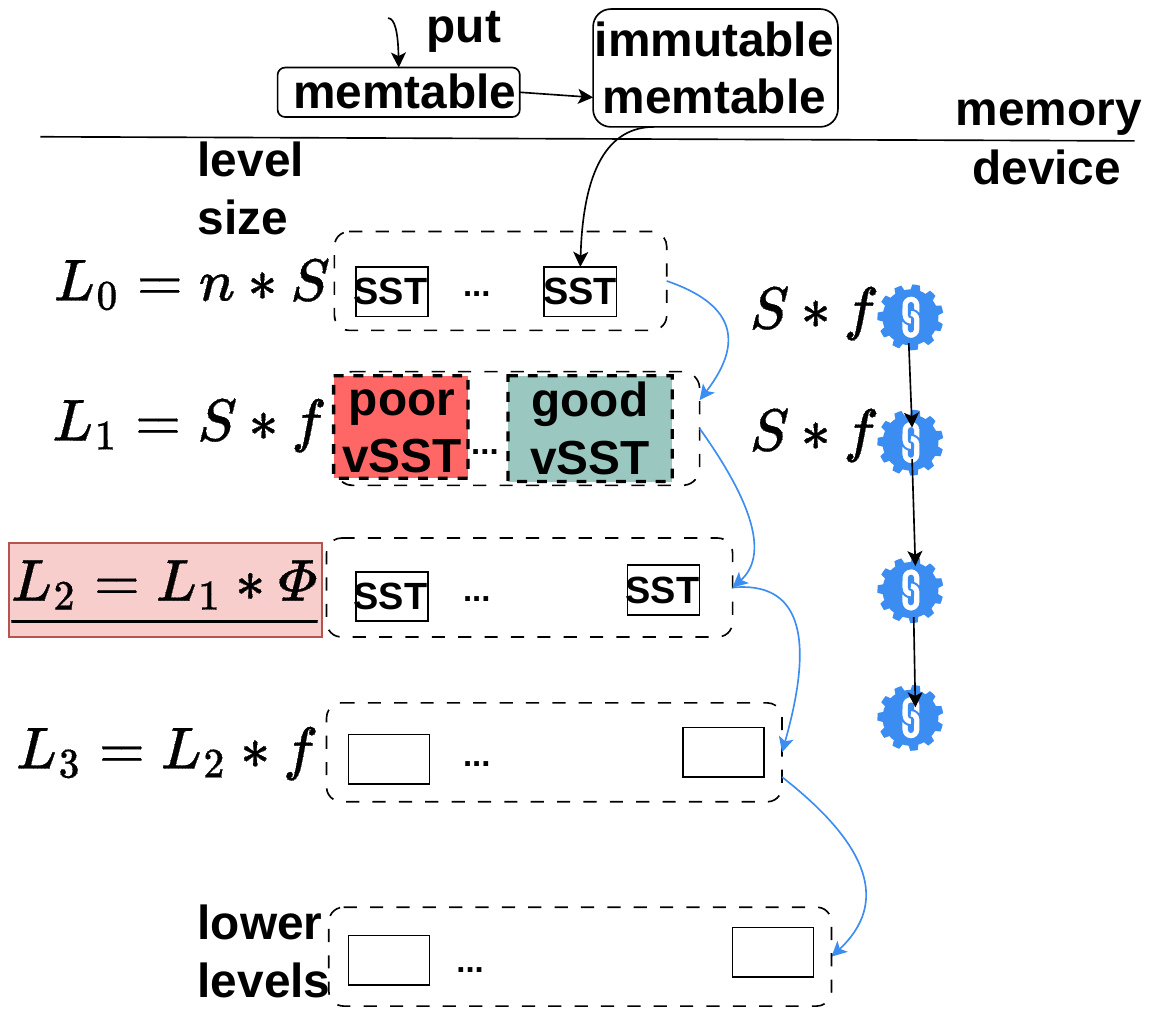}\label{fig:design}}
  \caption{ The main design points for mitigating high tail latency.}
\end{figure*}

\begin{figure}[t]
  \centering
  \subfigure[]{\includegraphics[width=0.49\columnwidth]{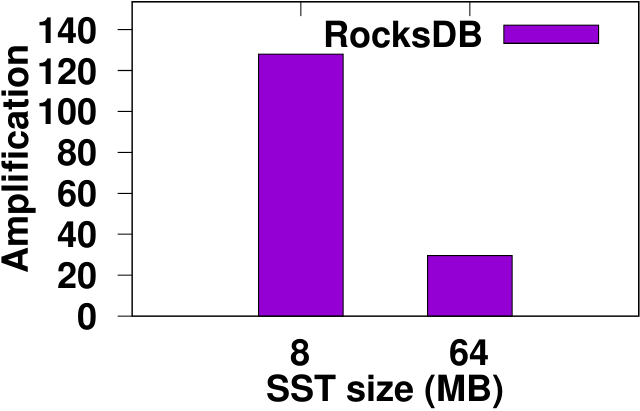}\label{fig:notiering-sstsize}}
  \subfigure[]{\includegraphics[width=.48\columnwidth]{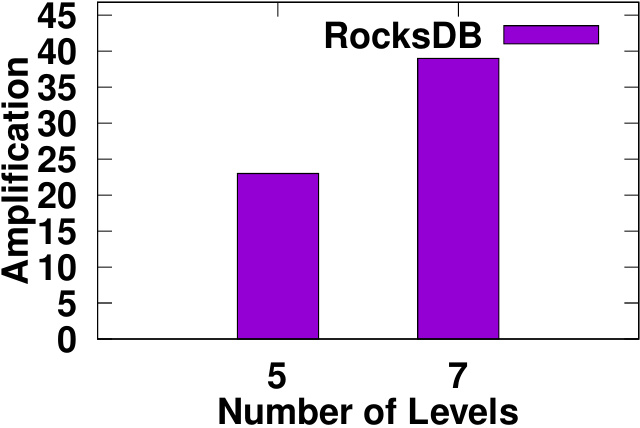}\label{fig:notiering-levels}}
  \caption{Impact on I/O amplification in RocksDB of (a) compacting a single SST between $L_0$ and $L_1$ when not maintaining
    growth factor between $L_0$ and $L_1$ and (b) the number of LSM levels when using 8 MB SSTs emulating LSMi.}
  \label{fig:ioampl}
\end{figure}

Problematically, with small SSTs, the number of levels in the LSM store increases, which in turn inflates I/O amplification.
Figure~\ref{fig:notiering-levels} shows that if we change the SST size in RocksDB from 64 MB to 8 MB and adjust $L_1$ size from 256 MB
to 64 MB to maintain the growth factor to f=8, the number of levels grows from 5 to 7 and I/O amplification increases by 69\%. As a
result, current systems cannot reduce SST size without significantly increasing I/O amplification.

Figure~\ref{fig:design-vanilla} illustrates the design of a typical LSM KV store with incremental compaction without tiering. If we try to
reduce the compaction chain width by reducing the SST size and removing the tiering compaction step at $L_0$, then I/O amplification
increases dramatically. Each SST in $L_0$ overlaps with the entire $L_1$, resulting in excessive I/O amplification.
Figure~\ref{fig:notiering-sstsize} shows that using the default configuration of RocksDB, after disabling the tiering compaction, the I/O
amplification increases up to 128$\times$, when $L_1$ is 256 MB in total. We reduce SST size from 64 MB to 8 MB, which effectively increases
the growth factor between $L_0$ and $L_1$ from f=4 to f=32. Therefore, current designs can only reduce the size of $L_1$ to be $f$ times the
SST size while increasing the number of levels, as we show in Figure~\ref{fig:notiering-levels}.

Finally, to reduce I/O in incremental compaction partially, operators often use over-provisioning of space as follows: They
keep each region at about 40\% of its maximum size~\cite{rocksdbtos}. Essentially, this keeps the last level relatively empty with
fewer SSTs.  When a compaction chain forms, the last stage incurs less work because the last-level compaction involves, on average, fewer than $f$ SSTs and reduces I/O amplification by up to 20\%~\cite{spooky}.

\subsection{Eliminating dependencies in compaction chains}
Recall that compaction chains are sequences of dependent compactions
from level to level.
We can use two alternative approaches to perform these compactions to reduce the delay until we can free memory for the blocked operation.

Modern LSM stores, including RocksDB, may allow levels to temporarily exceed their maximum size (compaction debt). In that case, they can perform compactions in the background, ignoring
the size limit for each level while serving regular requests. This approach results in a higher growth factor than $f$ across levels and, therefore, increases I/O amplification.
To avoid excessive I/O amplification, this approach places an upper per-level limit on the size of the compaction debt.
Zhou et al.~\cite{calcspar} expect to have compaction debt for short periods. In our evaluation, we show that allowing for compaction debt alone does not suffice to significantly reduce the tail latency.

On top of the compaction debt design, scheduling-based solutions also proactively perform compactions in the background. Figure~\ref{fig:design-adoc}, illustrates ADOC, a state-of-the-art scheduling-based solution, which extends RocksDB design with compaction debt.
In short, scheduling those compactions out-of-the-critical path, strives for insert requests to always have free space at each level, therefore preventing compaction chains from forming.
While scheduling compactions in the background of a running workload can drastically improve the tail latency, it comes at almost double the cost of I/O amplification (as shown in our evaluation~\cref{sec:eval}).


\subsection{Summary}
Overall, addressing the root causes of tail latency is a significant challenge in modern KV designs. Existing designs trade tail latency with I/O amplification or memory usage, and vice versa.
In many cases today to mitigate tail latency,
operators resort to circumstantial actions by either over-provisioning resources
or rate-limiting storage servers~\cite{fbookstudy} to avoid long queues of requests.
Next, we discuss how \name{} reduces tail latency without such compromises.

\section{\name{}}
\label{sec:design}

Unlike existing LSM works, \name{} manages to optimize for all three metrics: memory usage, tail latency, and I/O amplification.
To achieve that, \name{} carefully reduces both the maximum width and maximum height of compaction chains in LSMs by combining four key ideas. We discuss these ideas and their synergy in this section.

Figure~\ref{fig:design} illustrates the
write path of \name{}. Initially, \name{} writes KV pairs in the memtable in memory. When a memtable is full, \name{} serializes it
to an SST and flushes it to $L_0$. This step proceeds even if there are SSTs present from previous flush operations
until the size of $L_0$ reaches a predefined max number of SSTs, similar to RocksDB.  At this point, \name{} behaves differently from RocksDB, as we explain next.

\subsection{Reduce width: No tiering \& Smaller SSTs}

First, \name{} does not have to wait for $L_0$ to become entirely full as it does not have to perform a tiering compaction to $L_0$.
Instead, it picks a single SST from $L_0$, in FIFO order each time, and compacts it to $L_1$ to free space in $L_0$ for a memtable. As a
result, in \name{} $L_0$ serves merely as a queue of SSTs to handle traffic bursts efficiently. Although its size needs to be limited for
read performance purposes (as in RocksDB), it does not impose any limitations in reclaiming space with compactions due to eliminating the
tiering step (unlike RocksDB).

Second, contrary to RocksDB, \name{} always compacts a single SST from $L_0$ to $L_1$. To control I/O amplification, \name{}
uses by design an $L_1$ with size $f$ times the size of a single SST compared to RocksDB where $L_1$ size is equal to $L_0$.  As a result, the width of the compaction chain now depends
on the SST size and the growth factor, which affects the average overlap of SSTs between adjacent levels.

Naively removing the tiering step and decreasing the SST size to reduce the compaction width according to the SST size increases the number
of LSM levels, by a constant $f$ across all levels. As a result, \name{} would need to compact more levels for the same dataset size,
affecting both tail latency and I/O amplification. For example, using 64 MB SSTs with a 4 TB region capacity, \name{} would require two
additional levels at $f=8$.

Therefore, \name{}'s third design aspect is it uses a larger growth factor $\Phi$ between $L_1$ and $L_2$ compared to $f$ across the
rest of the levels, to maintain the number of levels the same as when using a tiering compaction.  Increasing the growth factor,
increases merge amplification~\cite{vat} between $L_1$ and $L_2$~\cite{vat} because it affects the amount of overlap across the two
levels.

To overcome this challenge, \name{} introduces its overlap-aware SSTs (vSSTs) technique in $L_1$ to limit merge amplification between
$L_1$ and $L_2$. \name{} creates in $L_1$ vSSTs of variable overlap, as follows.  When creating a vSST in $L_1$, \name{} examines the
overlap of the new vSST with $L_2$. Then, it allows some vSSTs to be smaller than regular SSTs with limited overlap to $L_2$. By design
\name{}, creates two types of vSSTs: \emph{poor} vSSTs with overlap more than $f$ and \emph{good} vSSTs with overlap less than or equal to
$f$. Then, during compaction from $L_1$ \name{} compacts only good vSSTs until it frees space for an SST in $L_1$ for the next $L_0$ SST. Next,
we discuss how vSSTs work in more detail.

\subsection{Reduce length: Larger growth factor $\Phi$ in $L_1$ \& overlap-aware SSTs}
To maintain the same number of levels as tiered approaches (RocksDB), \name{} uses a larger growth factor in $L_1$. Naively done, this would increase the
growth factor up to 32$\times$. To prevent the increase in the number of levels due to the smaller SSTs, as observed in the design of LSMi in
Figure~\ref{fig:design-vanilla}. The increased growth factor decreases the number of SSTs in $L_1$ and results in each $L_1$ SST to overlap
with a larger number of $L_2$ SSTs. Generally, the average overlap of SSTs between adjacent levels is $f$ in LSM KV stores. However, in
\name{} $L_1$ SSTs overlap with up to 32$\times$ more $L_2$ SSTs, resulting in a significant increase in I/O amplification. To overcome this
challenge, \name{} introduces overlap-aware SSTs in $L_1$ to control the overlap between $L_1$ and $L_2$ SSTs. Next, we discuss how vSSTs
work in more detail.


\name{} uses overlap-aware vSSTs only in $L_1$, while it keeps fixed-size SSTs in all other levels.  \name{} takes advantage of the following property: since each $L_0$ SST usually contains keys that cover the whole key range of $L_1$, it reorganizes the vSSTs in
$L_1$ on every $L_0$ compaction.

During each compaction from $L_0$ to $L_1$, before appending a key to the current in-flight $L_1$ vSST, it checks the overlap of
the vSST with the SSTs of $L_2$ as if the key is in the vSST. If the overlap exceeds $f$, it closes the current vSST, flushes
the vSST to $L_1$, and starts the next vSST. However, \emph{naively} closing a vSST when the overlap exceeds $f$ could lead to many
small vSSTs in $L_1$ (poor vSSTs), with high I/O amplification. Next, we discuss how \name{} creates a bounded number of vSSTs
while controlling I/O amplification for $L_1$ vSST.

\subsubsection{Good and poor vSSTs}

To prevent the excessive creation of poor vSSTs that lead to high I/O amplification due to the fixed-size SSTs of $L_2$, \name{}
uses a heuristic that limits the number of poor vSSTs in $L_1$. \name{} sets the minimum size of the vSSTs to $S_m = 1/f \times
      S_M$, where $S_M$ is the size of the fixed-size SSTs.

Essentially, \name{} needs to solve the following optimization problem in $L_1$. Given the keys in $L_1$ and the fixed-size SSTs in
$L_2$, we need to divide $L_1$ in vSSTs (vSSTs) with the following constraints.

Each vSST should have a size between a minimum ($S_m$) and a maximum ($S_M$) number of bytes. $S_M$ is the maximum size for a fixed-sized SST
in the system, e.g. 64 MBytes in the default RocksDB configuration and between 4-64 MB in \name{}. $S_m$ depends on the overall design of
the KV store. Generally, it should not be too small, e.g. in the order of KB, because this will result in a significant increase in per-SST
overheads, such as the size of memory guards (in the memory part of the index) and the cost of manifest flush operations in RocksDB. In our work,
we set $S_m = 1/f \times S_M$, which leads to at most $f$ more SSTs in $L_1$, compared to RocksDB's fixed-size SST approach.

Assuming each vSST has overlap $O$ (ratio of bytes between vSST and next-level SSTs) with the SSTs in $L_2$, \name{} needs to
divide $L_1$ in vSSTs in a manner, such that there is always \emph{enough} vSSTs with overlap $O$ less than f. If \name{} achieves
this, during the next compaction from $L_1$ to $L_2$, the system will pick a subset of these vSSTs for compaction, freeing adequate
space in $L_1$ for the next $L_0$ SST. As an approximation, instead of creating an adequate number of good vSSTs with appropriate
overlap, \name{} tries to maximize the number of such vSSTs. Therefore, we can state the objective as:

\begin{description}[leftmargin=*]
      \item[Objective:] Given the keys in $L_1$ and the fixed-size SSTs in $L_2$, divide $L_1$ in overlap-aware vSSTs with size
            between $S_m$ and $S_M$, such that you maximize the cumulative size of vSSTs that exhibit overlap $O$ less than f with $L_2$
            SSTs.
\end{description}

\name{} uses a heuristic approach to create vSSTs in $L_1$. First, it tracks the overlap $O$ of the next vSST with $L_2$. Then, it
keeps adding keys to the vSST until:
\begin{itemize}[leftmargin=*]
      \item If overlap $O$ becomes quickly larger than f, then it closes the vSST when it reaches size $S_m$. We call these
            \emph{poor} vSSTs as they could result in high compaction costs. \emph{Poor} vSSTs that have overlap more than f and their size
            is always m.
      \item If overlap $O$ is less than f, it keeps appending to the vSST until either the overlap becomes f or its size reaches $S_M$. We
            call these \emph{good} vSSTs as they result in low compaction cost. \emph{Good} vSSTs have overlap up to f, and their size by
            necessity is between [m,M].
\end{itemize}

During the next incremental compaction from $L_1$ to $L_2$, \name{} picks as many \emph{good} vSSTs as necessary to free adequate space in
$L_1$. Although, in principle, it is possible that there are not enough \emph{good} vSSTs in $L_1$, due to the growth factor F between $L_1$
and $L_2$, which is larger than f, our evaluation shows that for values of $\Phi$ up to 32, the system is able to always find \emph{good} vSSTs.
By design, \name{} allows creating \emph{poor} vSSTs that overlap (usually larger than $f$) with a large part of $L_2$. These \emph{poor}
vSSTs will not be used during the next compaction, but they will be reorganized later when a new $L_0$ SST is compacted to $L_1.$ At that
point \name{} will try to create again \emph{good} vSSTs, even though it is starting from \emph{poor} vSSTs in $L_1$. \name{} takes
advantage of \emph{poor} vSSTs by making them have really large overlap with $L_2$ SSTs and this forces the remaining \emph{good} vSSTs to
have lower overlap less than $f$. Also, \name{} removes the constraint of fixed-size SSTs in $L_1$ to free space for the next $L_0$ SST by
allowing the compaction of multiple \emph{good} vSSTs to $L_2$ without increasing I/O amplification.

Given the value for $S_m = 1/f \times S_M$, \name{} creates at most $f$ times more vSSTs in $L_1$ than fixed-size SSTs (each of
size $S_M$) it would have. We could reduce this number further with a larger value for $S_m$. However, increasing $S_m$ has a
detrimental effect: If we keep growing a \emph{poor} vSST, this reduces the opportunity for identifying low-cost vSSTs by
\emph{absorbing} a key range that on its own could form a \emph{good} vSST. Generally, other values for $S_m,$ we speculate smaller
than the value we currently use, could be appropriate for different KV store designs, e.g, with lower constant overheads compared
to RocksDB. Next, we discuss how \name{} selects \emph{good} vSSTs.

\subsubsection{Selecting \emph{good} vSSTs to compact}
\name{} uses the same compaction scheduler as RocksDB to select the best \emph{good} vSSTs in the following manner. The default compaction
scheduler of RocksDB first randomly selects 50 SSTs from the source level ($L_1$ in our case). Due to the size of $L_1$, it typically
examines all its SSTs. Then, it calculates for each vSST the ratio $\frac{overlap\_in\_bytes\_of L_2}{SST\_size\_of L_1}$. As a result, it
chooses the largest vSST(s) of $L_1$ in size with less overlap with $L_2$. Finally, it compacts a set of SSTs from $L_1$ whose cumulative
size equals the fixed SST size ($S_M$).


\section{Experimental Methodology}
\label{sec:method}

\beginbsec{Experimental platform}
Our testbed consists of a single server that runs the key-value store and YCSB ~\cite{YCSBC} or db\_bench. Our server is equipped with two
Intel(R) Xeon(R) CPU E5-2630 running at 2.4~GHz, with 16 physical cores for a total of 32 hyper-threads and with 256~GB of DDR4
DRAM. It runs CentOS 7.3 with Linux kernel 3.10.0. The server has 1 Samsung SSD 970 EVO Plus 2TB device model.

\beginbsec{Workloads}
In our evaluation, we use YCSB~\cite{YCSBC} and db\_bench~\cite{dbbench}. We configure YCSB to use 15 client threads with four regions, and
each shard uses four threads for background I/O operations (compactions), on top of \textit{XFS} with disabled compression and
jemalloc~\cite{jemalloc}, as recommended. We configure RocksDB to use direct I/O, as recommended by RocksDB~\cite{rocksdbdio}, and we verify
that this results in better performance in our testbed than using the kernel page cache. We set the size of $L_1$ to 256~MB, the growth
factor to 8, and the maximum number of LSM levels to 5. We use YCSB to benchmark RocksDB and \name{} with 350 million key-value pairs for
Load A and 70 M operations for Run A-D workloads. We set the size of each key-value pair to 200~B. Finally, we configure all KV stores
overflow limit to 64 MB for all configurations to accelerate all LSM stores to reach a stable state. We set the default memtable and SST
size to 64 MB for RocksDB, unless stated otherwise.

We measure tail latency of write operations with YCSB Load A under the default uniform key distribution. We choose this
workload similar to previous studies~\cite{silk,matrixkv2} because it represents write-intensive workloads.
We configure the different baselines throughout this study, following the general guidelines in~\cite{rocksdb} and introducing the required
parameter modifications, as discussed for each case. Furthermore, we use YCSB Run A, B, and D, which contain different mixes of read and
write operations to evaluate the performance of the different systems under mixed workloads. Moreover, to understand how \name{} affects
mixed workloads, we extend the YCSB completion queue to differentiate between read and write operations. For Runs A, B, and D, we present a
breakdown of the tail latency of read and write operations separately, e.g. for Run A we indicate write and read latency as Run A-W and Run
A-R, respectively.

We use db\_bench to evaluate production workloads as in Meta's datacenters~\cite{fbookstudy}. We use the same configuration noted in
~\cite{fbookstudy} for the KV store population but increase the number of KVs from 50 million to 908 million while maintaining the remaining
parameters the same. We increase the number of KVs to fill all the LSM levels except the last one and ensure we measure the system in a
steady state. We measure with db\_bench only I/O amplification as the modifications needed to measure tail latency (as we do with YCSB)
require significant changes that might change the benchmark's behavior. However, we do not expect tail latency to differ substantially from
the results we report with YCSB.
\begin{figure}
  \includegraphics[width=\columnwidth]{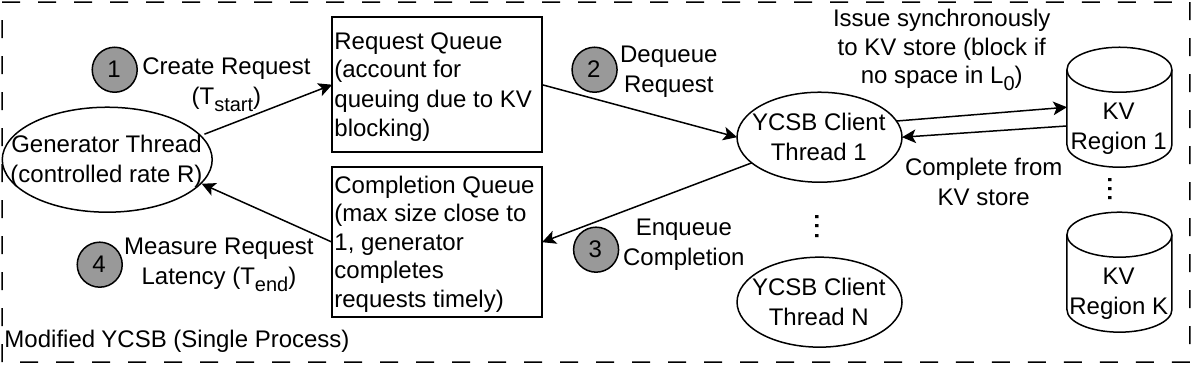}
  \caption{Modified YCSB to measure tail latency in an open-loop manner at a controlled (fixed) request rate.}
  \label{fig:YCSB}
\end{figure}

\begin{figure*}[t]
  \subfigure[Latency]{\includegraphics[width=.33\textwidth]{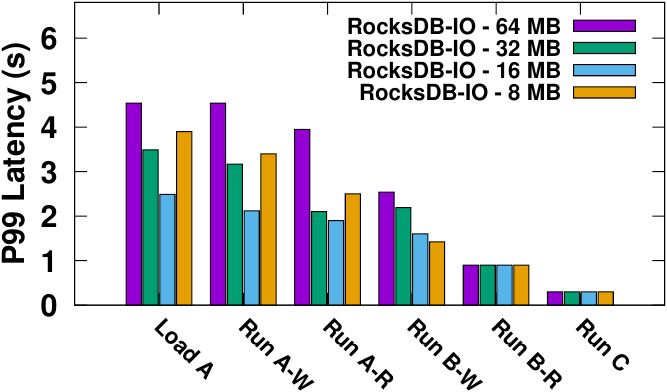}\label{fig:full-ycsb-rdblat}}
  \subfigure[Throughput]{\includegraphics[width=.33\textwidth]{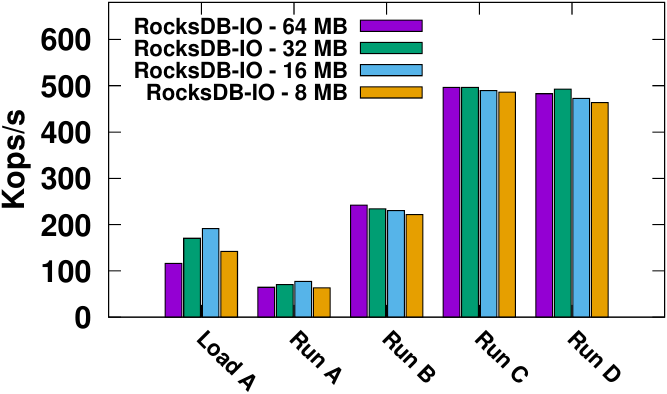}\label{fig:full-ycsb-rdbthroughput}}
  \subfigure[CPU Efficiency]{\includegraphics[width=.33\textwidth]{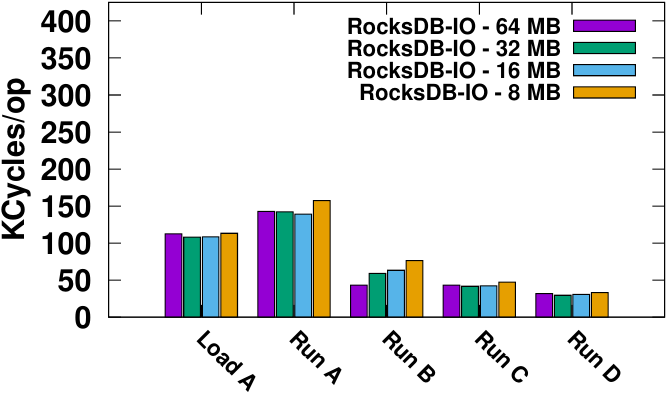}\label{fig:full-ycsb-rdbcycles}}
  \caption{RocksDB P99 tail latency
    (left), throughput (middle), and CPU efficiency(right) for all YCSB workloads while varying
    the SST size between 8-64 MB.}
  \label{fig:full-rdbycsb}
\end{figure*}

\beginbsec{Tail latency}
Similar to previous work~\cite{silk,matrixkv2,adoc}, we use two metrics to study tail latency: a) \emph{write stalls} and b)
\emph{P99 latency} as seen by the application. Write stall metric is the time server throughput drops to zero due to memory
unavailability at $L_0$. They offer an indirect view of operation completion times.

There are two methods to measure tail latency in a system: a) Each client issues a number of requests and stops generating new
requests until it gets a reply from the server. b) Requests are sent at a constant rate regardless of the timing of the responses
received from the server.

Case (a) leads to coordinated omission scenario~\cite{coordom}, where the request rate is inadvertently tied to the server's
ability to process requests, skewing tail latency measurements~\cite{tailbench,treadmill} due to reduced queueing.
In addition, in cloud environments with many independent clients, even if individual clients bound their requests to the KV store,
the cumulative number of requests from all clients cannot be controlled. Therefore, in our methodology, we use (b), which reflects
the tail latency seen by applications. To examine different configuration points, we bound in each experiment the \emph{rate} of
the requests.


We modify YCSB~\cite{YCSB,YCSBC} to measure tail latency, as shown in Figure~\ref{fig:YCSB}. We first decouple the generation of KV
pairs from client threads by introducing a new thread that generates requests and places them in an unbounded queue at a fixed
request rate (\circled{1}). Each request includes the operation type, key, and timestamp. Several YCSB threads dequeue requests
from the unbounded queue and issue them to the KV store \emph{synchronously} (\circled{2}) via synchronous operations.
When each request completes, the corresponding YCSB thread detects completion and moves the request to a completion queue
(\circled{3}). All YCSB threads share the request completion queue. The generator thread dequeues completed requests and measures
end-to-end, per-request tail latency using the request issue timestamp (\circled{4}). We ensure accuracy by using the same core
clock counter for end-to-end measurements, avoiding discrepancies caused by unsynchronized core clocks.

In our approach, the single generator thread limits the maximum request generation and completion rate. We find that a single
thread can generate and complete requests at a maximum rate of 1.5M requests/s, significantly higher than the maximum throughput of
RocksDB and \name{}, which is approximately 200K request/s.

Additionally, we use the following technique to set the appropriate request rate without inducing excessive queuing effects. We
differentiate between maximum throughput and sustainable throughput.  We conduct profiling runs setting the generator to operate at a large
rate and we identify the sustainable throughput for each system. Previous works~\cite{silk,matrixkv2} measure tail latency at low request
rates. In this paper we measure tail latency under broad load conditions, up to 95\% of its sustainable throughput, which provides a clearer
understanding of the impact of each design on tail latency.

\beginbsec{CPU efficiency}
Finally, we measure CPU efficiency of the KV store in cycles/op. \mathvar{cpu\_util\%} (in the range [0,1]) is the average of CPU
utilization among all processors, excluding idle and I/O wait time, as given by~\textit{mpstat}. As $cycles/s$ we use the per-core clock
frequency. Finally, $\mathvar{average\_ops}/s$ is the throughput reported by YCSB, and $\#cores$ is the number of system cores, including
hyper-threads.

\hspace{50pt}$cycles/op = \frac{cpu\_util~\times~cycles/s~\times~\#cores}{average\_ops/s}$

\beginbsec{Baselines} In our evaluation, we use as baselines the design variations discussed during our design: RocksDB,
RocksDB-IO, ADOC and \name{}. RocksDB is the default configuration of RocksDB. RocksDB-IO is a variant of RocksDB with overflow
(debt) disabled. We use RocksDB-IO to evaluate the effects of tiering compaction and how it affects tail latency without the
mitigating impact of debt which also increases I/O amplification. We use ADOC as the state-of-the-art scheduling approach that
optimizes tail latency while also reducing I/O amplification compared to RocksDB (but has worse I/O amplification than RocksDB-IO).

\section{Experimental Evaluation}
\label{sec:eval}

To understand the impact of each design on tail latency and how effective \name{} is, we examine the following questions:
\begin{enumerate}[leftmargin=*]
      \item How does \name{} perform compared to RocksDB and ADOC regarding tail latency, throughput, I/O amplification, and CPU efficiency?
      \item How does \name{} reduce compaction chain length \& width?
      \item How does \name{} behave for different growth factors $\Phi$ and SST sizes?
\end{enumerate}

Next, we examine each of these aspects.

\begin{figure}[t]
      \subfigure[Write stalls]{\includegraphics[width=.33\columnwidth]{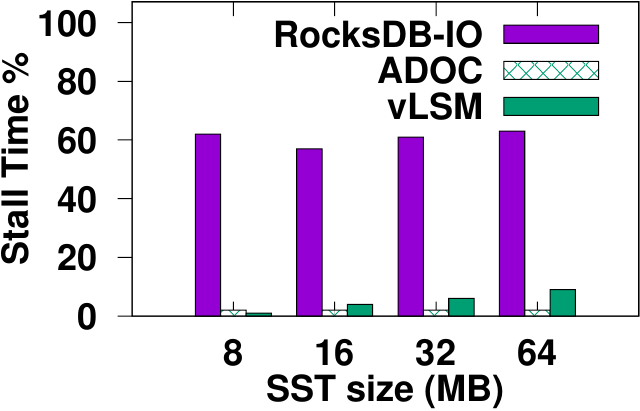}\label{fig:var-sst-size-stallsrdb}}
      \subfigure[Max stall]{\includegraphics[width=.32\columnwidth]{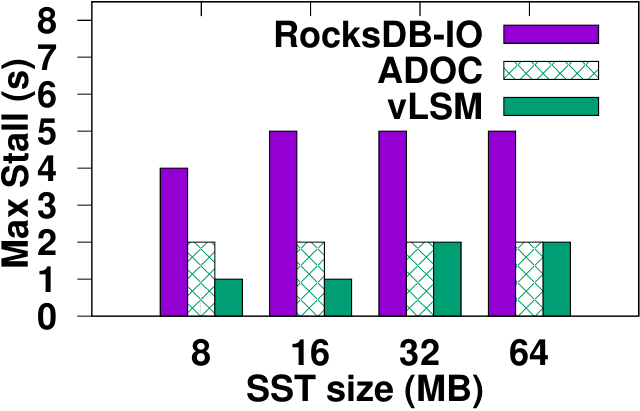}\label{fig:var-sst-size-max-stallsrdb}}
      \subfigure[I/O amplification]{\includegraphics[width=.33\columnwidth]{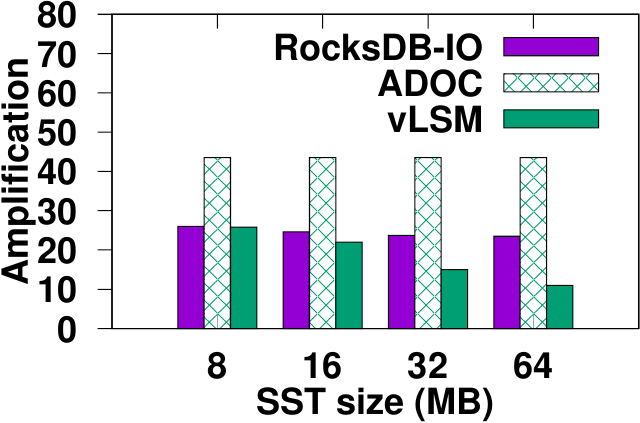}\label{fig:var-sst-size-amprdb}}
      \caption{RocksDB, ADOC and \name{} write stalls (left), max stall time (middle), and I/O amplification (right) for YCSB's Load A while varying the
            SST size between 8-64 MB.}
      \label{fig:var-sst-sizerdb}
\end{figure}

\subsection{Compaction chains and tail latency in the state-of-the-art}
First, we examine the impact of SST size on tail latency in RocksDB-IO and ADOC~\cite{adoc} to establish the context for modern,
state-of-the-art KV stores (Figure~\ref{fig:full-rdbycsb}). We run YCSB workloads Load A and Run A-D and report tail latency,
throughput, and CPU efficiency using four regions and four client threads. Figure~\ref{fig:full-ycsb-rdblat} shows tail latency for
RocksDB decreases compared to 64 MB SST size to 1.4$\times$ and 1.8$\times$, respectively, for 32 and 16 MB SST size. Then, for 8
MB, it increases again, as in the 32 MB SST size case.

Furthermore, SST size does not affect I/O amplification and stall time, as shown in Figure~\ref{fig:var-sst-size-stallsrdb}. The
compaction chains formed in RocksDB consist of two stages. The first stage, includes the tiered $L_0$, and $L_1$ compactions, whose
width is unaffected by the SST size. However, SST size affects the width of the chain of the lower levels. As a result, we observe
the drop in latency by up 1.8$\times$; however, tail latency for RocksDB remains in the order of seconds. We observe write stalls
in RocksDB for all SST sizes, as shown in Figure~\ref{fig:var-sst-size-stallsrdb} and~\ref{fig:var-sst-size-max-stallsrdb}. The
write stalls remain constant for all SST sizes, and the maximum stall time is in the order of 4-5 seconds. This is an effect of
tiering compaction in RocksDB, which compacts 2.5 GB of data from $L_0$ on average. Although ADOC increases the chain width due to
the combination of tiering compaction and the level overflow it dramatically reduces stalls similarly to \name{}. However, ADOC
increases I/O amplification by 1.8$\times$ compared to RocksDB and \name{}.

\begin{figure}[t]
      \centering
      \includegraphics[width=0.35\textwidth]{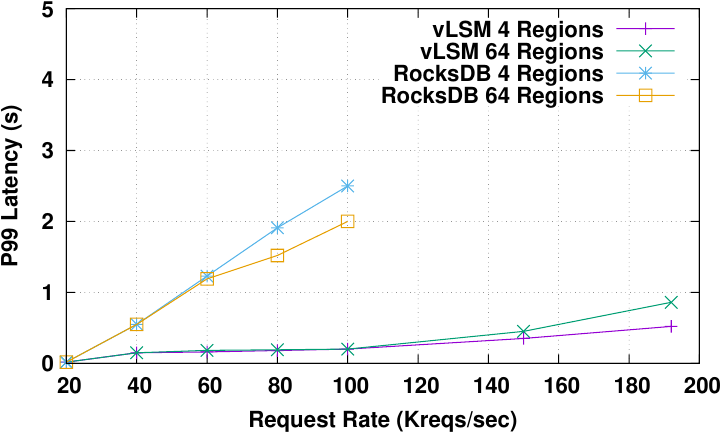}
      \caption{P99 latency for \name{} and RocksDB at different request rates.
      }
      \label{fig:reqratelat}
\end{figure}

\subsection{Compaction chains and tail latency in \name{}}
Next, we examine how \name{} behaves with respect to compaction chain length and width and how different workloads affect tail
latency and write stalls.

\beginbsec{Request rate}
Figure~\ref{fig:reqratelat} shows the P99 latency for RocksDB and \name{} with 8 MB SSTs and different
request rates. We stop measuring RocksDB after 100K requests/s as this is our setup's maximum sustainable throughput for
RocksDB. We find that when RocksDB serves more than 60K requests/s, tail latency exceeds 1 second while \name{}'s tail latency
remains below 1 second for all request rates. We note that the request rate does not significantly affect the tail latency for
RocksDB since the amount of work done in $L_0$ is large for all request rates due to tiering compaction. In contrast, \name{}'s
tail latency is not affected so much by the request rate since the amount of work done in $L_0$ is reduced due to the reduced
compaction chain length and width. For four regions, \name{} achieves tail latency up to 4.3$\times$ and for 64 regions up to
4$\times$ less than RocksDB.

\begin{figure}
      \subfigure[Chain Width] {\includegraphics[width=0.49\columnwidth]{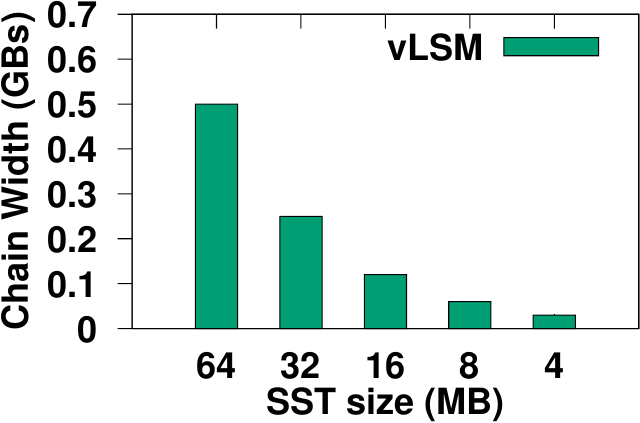}\label{fig:cc-width}}
      \subfigure[Chain Length] {\includegraphics[width=0.50\columnwidth]{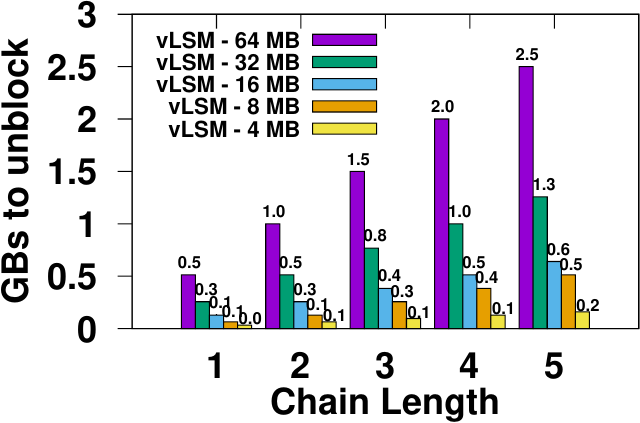}\label{fig:cc-length}}
      \caption{\name{} chain width (left) and chain length (right) for different SST sizes.}
\end{figure}

\begin{figure}[t]
      \subfigure[Latency]{\includegraphics[width=.49\columnwidth]{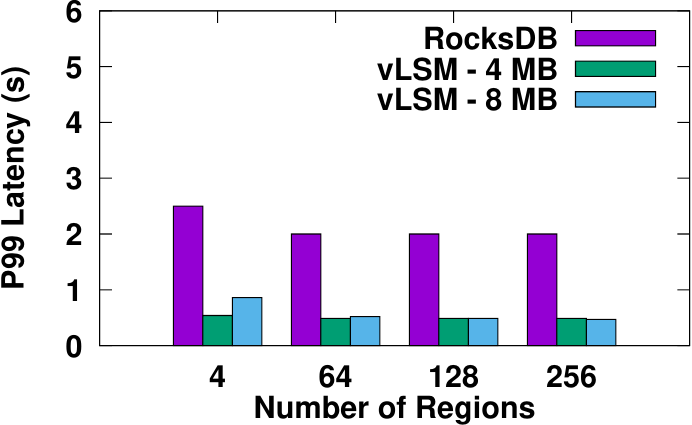}\label{fig:shardslat}}
      \subfigure[Throughput]{\includegraphics[width=.49\columnwidth]{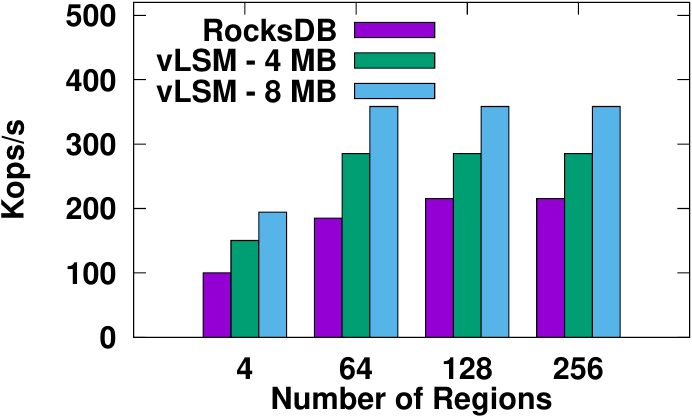}\label{fig:shardsxput}}
      \caption{Impact of \name{} on tail latency and throughput under YCSB Load A and varying number of regions.}
      \label{fig:shards}
\end{figure}

\beginbsec{Compaction chain width}
Figure~\ref{fig:cc-width} shows the compaction chain width with different SST sizes for \name{}. For \name{} chain width decreases as the
SST size up to 320$\times$ compared to RocksDB from 10 GB to 32 MB for 4 MB SSTs. The improvement in the chain width is due to the
removal of tiering compaction and the reduction of the SST size.

\begin{figure}[t]
      \centering
      \includegraphics[width=0.80\linewidth]{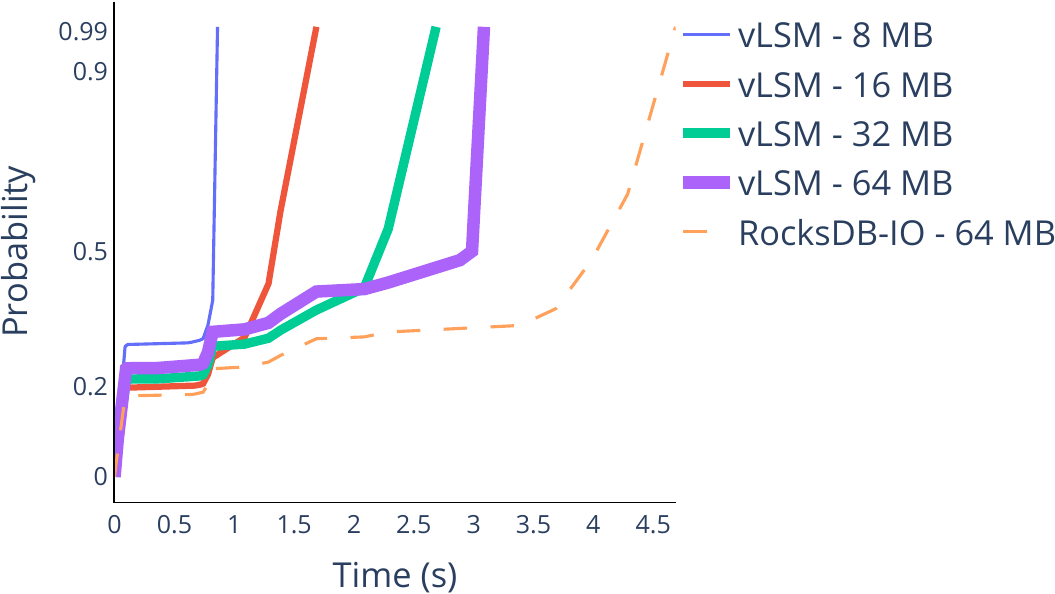}
      \caption{Open loop YCSB Load A tail latency CDF for RocksDB and \name{}.}
      \label{fig:cdfloada}
\end{figure}

To measure the impact of the compaction chain width in write stalls and tail latency, we use YCSB Load A and vary the SST size between 8-64
MB, while maintaning the number of regions at 4 in Figure~\ref{fig:var-sst-size-stallsrdb}. We see that \name{} reduces write stalls up to
60\% compared to RocksDB-IO when comparing \name{}'s 8 MB SSTs with RocksDB-IO as \name{} reduces the amount of data compacted from $L_0$ to $L_n$
51$\times$ compared to RocksDB-IO. Additionally, in Figure~\ref{fig:var-sst-size-max-stallsrdb}, we observe that \name{} reduces the maximum
stall time by 5$\times$ compared to RocksDB-IO. We note that the maximum stall time for RocksDB-IO is in the order of seconds for all SST sizes,
while for \name{}, it is in the order of a second. Additionally, we note that \name{} does not only improve P99 latency but also P50 and
P90. In Figure~\ref{fig:cdfloada}, we measure different percentiles of RocksDB-IO with 64 MB SSTs and \name{} with 8 to 64 MB SSTs for YCSB
Load A. We observe that \name{} for 8 MB SSTs improves P50 up to 4$\times$ and P90 up to 4.88$\times$ compared to RocksDB-IO.

\begin{figure*}[t]
      \subfigure[Latency]{\includegraphics[width=.30\textwidth]{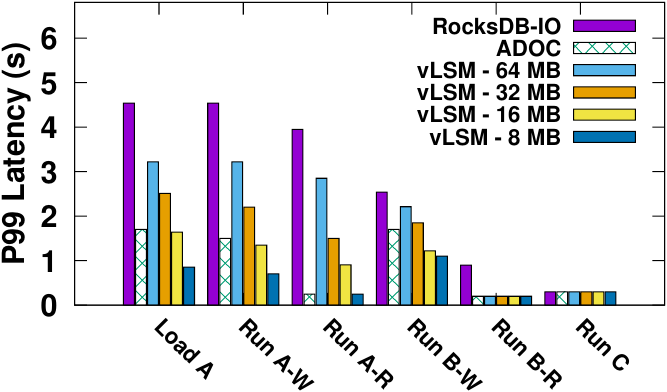}\label{fig:full-ycsb-lat}}\hfill
      \subfigure[Throughput]{\includegraphics[width=.30\textwidth]{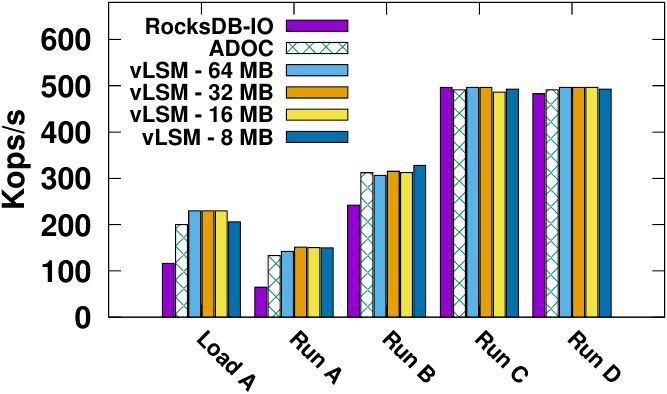}\label{fig:full-ycsb-throughput}}\hfill
      \subfigure[CPU Efficiency]{\includegraphics[width=.30\textwidth]{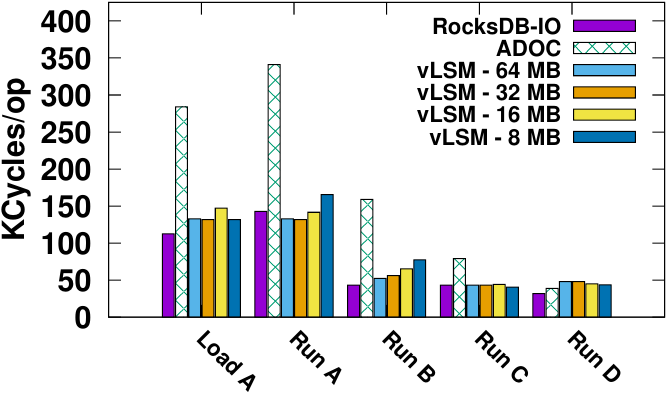}\label{fig:full-ycsb-cycles}}

      \caption{P99 latency (left), throughput (middle), and CPU efficiency (right) of \name{} for all YCSB workloads while
            varying the SST size between 8-64 MB. }
      \label{fig:full-ycsb}
\end{figure*}

\beginbsec{Compaction chain length}%
To examine the impact of \name{} on compaction chain length, we measure the amount of data (GBs) compacted before a memtable can be
flushed to $L_0$. We vary the number of levels from 1 to 5, which corresponds to different chain lengths. In
Figure~\ref{fig:cc-length}, we find that the amount of data in \name{} decreases up to 20$\times$ compared to RocksDB from 10 GB to
0.5 GB for 8 MB SSTs.

To measure the impact of the compaction chain length on tail latency, we vary the number of regions ranging from 4 to 256 in
Figure~\ref{fig:shardslat}. The number of regions affects the number of levels in the LSM tree since, for more regions, the number of levels
and the compaction chain length decrease. For our dataset size, with four regions, the LSM tree has five levels, and the compaction chain
length is four, while for 16 regions and above, each shard has three levels, and the compaction chain length is 3. We find that RocksDB,
even with sharding, has tail latency in the order of seconds for all shard counts and after 16 regions, where tail latency improves by
1.4$\times$ compared to four regions, it remains constantly high. This stems from the fact that the amount of work done from RocksDB comes
from tiering compaction in $L_0$, which on average is 2.5 GBs for five levels and 1.2 GB for three LSM levels. Compared to \name{}, which
compacts on average 0.32 GB for five levels and 0.19 GB for three LSM levels for 8 MB SSTs and 0.16 GB for five levels and 0.12 GB for 4 LSM
levels for 4 MB SSTs. \name{} achieves 4.23$\times$ better tail latency than RocksDB with 8 MB SSTs and 5$\times$ better tail latency with 4
MB SSTs for four regions. For 16 regions and above, \name{} with 4 MB SSTs improves tail latency by 6\% compared to four regions since for
16 regions 4 MB SSTs require 1 more level in the LSM tree to maintain $\Phi$ below 64.

Figure~\ref{fig:shardsxput} shows that \name{}, improves throughput for all SST sizes up to 9$\times$. Furthermore, for 4 MB SSTs \name{}
achieves 1.76$\times$ better latency in the expense of 20\% more I/O amplification since we need to add one more level to maintain the 32
ratio between $L_1$ and $L_2$. We also note that \name{} uses the same amount of memory for the in-memory component as the base RocksDB
configuration 2 memtables (as suggested by RocksDB) equal to the SST size.

\subsection{Sensitivity analysis for \name{}}

\beginbsec{Sensitivity to the workload}
Next, we compare the impact of the SST size on all YCSB workloads. We focus especially on mixed workloads, Run A and Run B that
have a mix of read and write operations to understand how writes affect read tail latency. We measure throughput, P99 Latency, and
CPU efficiency.

In Figure~\ref{fig:full-ycsb}, we run all YCSB workloads for Run A that has 50\% read and 50\% update operations \name{} achieves up to
2$\times{}$ better throughput since update operations either complete faster on higher levels or due to the reduced SST size that requires
less work. We note here that the tail latency for reads in Run A also improves up to 12.5$\times{}$ for 8 MB SSTs compared to RocksDB-IO since
read operations have to wait for significantly less time due to the large reduction in write stalls. Furthermore, we observe the same trend
for Run B, which has 95\% read and 5\% update operations.

For Run C that has 100\% read operations \name{} achieves the same throughput as RocksDB-IO, as SST size does not affect how an LSM tree
handles read operations. Only the block cache and data block sizes can significantly affect reads. Finally, we measure CPU efficiency and
observe that for all workloads containing write operations, \name{} decreases CPU efficiency, as it has to check for every KV pair, the
overlap with the next-level SSTs, and perform compactions more often. \name{} decreases CPU efficiency up to 4\% compared to RocksDB-IO. We
note that ADOC for write heavy workloads (Load A) and mixed workloads (Run A) decreases CPU efficiency up to 2.2$\times$ due to the overflow
mechanism that increases the average compaction size by 1.8$\times$. Furthermore, due to the overflow mechanism, ADOC decreases CPU
efficiency for Run C up to 2$\times$ since it has to perform compactions to reduce the debt from the previous write heavy workloads.

\begin{figure}
      \subfigure[Growth factor $\Phi$] {\includegraphics[width=0.33\columnwidth]{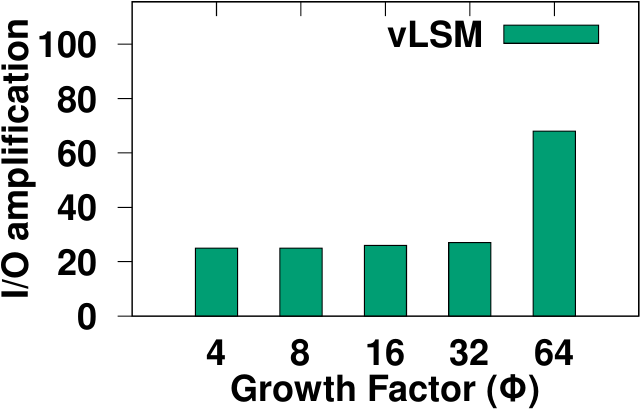}\label{fig:ampsizeratio}}
      \subfigure[File overlap] {\includegraphics[width=0.33\columnwidth]{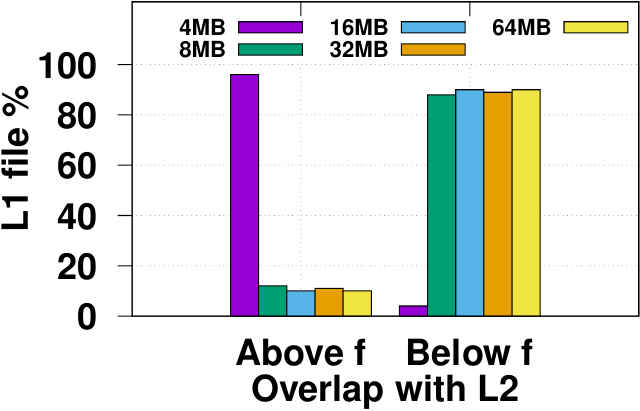}\label{fig:L1fileoverlap}}
      \subfigure[KV distribution]{\includegraphics[width=0.32\columnwidth]{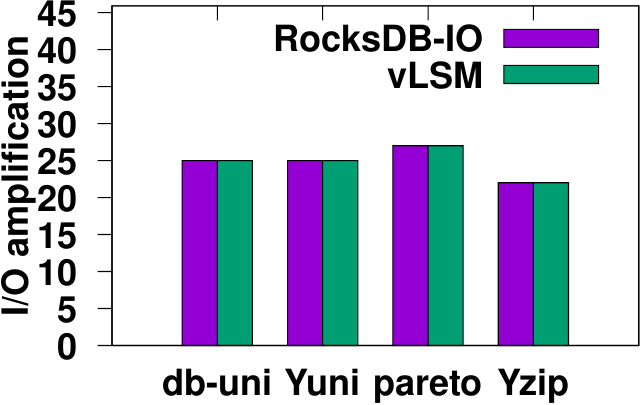}\label{fig:diffworkloads}}
      \label{fig:L1sizeranges}
      \caption{Sensitivity analysis for different vaules of $\Phi$, showing I/O amplification (left) and percentage of vSST overlap
            (middle). Sensitivity analysis with different key distributions (right).}
\end{figure}

\beginbsec{Sensitivity to the KV distribution}
In this experiment, we investigate \name{} vSSTs behavior for different workloads observed in large-scale applications. We run two
benchmarks with different key distributions: a) YCSB with uniform and Zipfian and b) db\_bench with uniform and Pareto. Pareto distribution
reflects Meta's workloads~\cite{fbookstudy}. Also, both uniform key distributions of YCSB and db\_bench generate a high-entropy workload
typically seen in large-scale applications. We measure \name{} with 8 MB and RocksDB-IO with 64 MB SSTs, respectively. We set the growth
factor between $L_1$ and $L_2$ to 32 for \name{}. In Figure~\ref{fig:diffworkloads}, we find that \name{} achieves the same I/O
amplification with RocksDB-IO for all distributions. We conclude that variable size SSTs of $L_1$ keep I/O amplification equal to RocksDB-IO
for the aforementioned KV distributions.

\beginbsec{Sensitivity to $\Phi$}
In \name{}, the role of the growth factor $\Phi$ from $L_1$ to $L_2$ is to reduce the number of LSM levels, therefore \name{} uses larger
values than what is typical in LSM (e.g. $f=8-10$). We examine different values for $\Phi$ by varying the SST size between 64 - 4 MB. By
varying the SST size, we affect the compaction size between $L_0$ and $L_1$. Varying the SST size directly affects the $L_1$ size since
compactions from $L_0$ to $L_1$ must maintain the growth factor $f$.

In Figure~\ref{fig:ampsizeratio}, we find that \name{} cannot sustain I/O amplification below $f$ when the $L_2/L_1$ ratio reaches 64 (4 MB
SSTs). To verify why \name{} cannot sustain I/O amplification below $f$ when the $L_2/L_1$ ratio reaches 64 for 4 MB SSTs, we analyze the
file sizes for 8 MB and 4 MB SSTs $L_1$. We measure the file sizes created on $L_1$ and separate the files into two categories. The first
category contains files that are exactly on the boundary of $1/f$ that exceeds $f$ overlap and the remaining files above $1/f$ that do not
exceed $f$ overlap. In Figure~\ref{fig:L1fileoverlap}, we find that for 8 MB SSTs, 90\% of the files do not exceed $1/f$ and 10\% exceeds
overlap $f$ across all compactions. As a result, every $L_1$ compaction finds a file that does not exceed $f$ overlap. Thus, \name{} can
sustain I/O amplification below $f$ for 8 MB SSTs. In contrast, for 4 MB SSTs, 94\% of the files exceed $f$ overlap, and 94 \% of the files
are exactly on the $1/f$ boundary.  On average, the work per byte for 4 MB SSTs is 8192$\times$ more than 8 MB SSTs when the $L_2/L_1$ ratio
becomes 64.

\beginbsec{Sensitivity to SST size}
Smaller SSTs have the potential to reduce tail latency by reducing the amount of work in compaction chains. Although modern storage devices
exhibit high IOPS and are not as sensitive to SST size, small SSTs increase the number of SSTs, which can affect other overheads. Especially
in modern KV stores that are designed for large SSTs, certain aspects of the design do not necessarily cater to small SSTs. For instance,
RocksDB manages recovery by flushing a system manifest to the device after every memtable flush and SST compaction. Small SSTs will lead to
more manifest flush operations. Similar issues exist with guard management in-memory and the layout of internal SST metadata.

\begin{table}[t]
      \begin{tabular}{|r|r|r|r|}
            \hline
            \begin{tabular}[c]{@{}c@{}}SST size\\ (MB)\end{tabular}     &
            \begin{tabular}[c]{@{}c@{}}P99 lat \\ (ms)\end{tabular}     &
            \begin{tabular}[c]{@{}c@{}}xput    \\ (Kops/s)\end{tabular} &
            \begin{tabular}[c]{@{}c@{}}CPU     \\ (Kcycles/op)\end{tabular}               \\ \hline
            2                                                           & 421 & 39  & 320 \\ \hline
            4                                                           & 502 & 161 & 196 \\ \hline
            8                                                           & 860 & 194 & 140 \\ \hline
      \end{tabular}
      \caption{Sensitivity analysis of \name{} based on SST size.}
      \label{tab:sst-limit}
\end{table}

We explore the effect of smaller SSTs in \name{}, focusing on throughput, latency, and CPU efficiency in Table~\ref{tab:sst-limit}. We find
that \name{}'s performance drops below 8 MB SSTs when using SST sizes between 2-4 MB. Our findings reveal a 15\% decrease in throughput and
CPU efficiency at 4 MB due to frequent compactions and overlap checks for each KV pair despite a 58\% improvement in P99 latency. Moving to
2 MB SSTs further reduces throughput by fivefold and CPU efficiency by 2.38$\times{}$, with only a 15\% latency improvement over 4 MB SSTs.
Furthermore, the increased CPU overhead as the SST size decreases stems from RocksDB's current design. Specifically, RocksDB is optimized to
handle larger SSTs and synchronously issues all the I/O path. Future LSM-tree designs could reduce CPU overhead by batching I/O operations
and compactions and reach for far smaller SST sizes with better performance compared to \name{}. \name{}'s performance is optimal with 8 MB
SSTs, as it achieves the best balance between throughput, tail latency, and CPU efficiency.

\section{Related Work}
\label{sec:related}

We group related work that directly targets improving tail latency in KV stores in the following three categories: (a) reduce the width of
the compaction chains, and (b) improve concurrency through I/O scheduling approaches.

\beginbsec{Compaction chain reduction}
MatrixKV~\cite{matrixkv2} is an LSM KV store that removes the tiering step between $L_0$ and $L_1$. It increases the size of
$L_0$ to reduce the number of LSM levels and dynamically creates SSTs that compact with $L_1$. As a result, it reduces both I/O
amplification and tail latency. NoveLSM~\cite{novelsm} adds an extra level between memtables and storage to store immutable memtables in
NVM. This allows NoveLSM to have a large temporary buffer (NVM) before compacting memtables to the storage. ChameleonDB~\cite{chameleondb}
uses lazy leveling with NVM for storage. This reduces I/O amplification significantly since it uses tiering for the N-1 levels except the
last. However, when the N-1 levels become full, the system will stall for a large amount of time since it will need to compact each full
level to the next thus observing long tail latency.

MioDB~\cite{miodb} uses NVM to store skiplists instead of SSTs for the N-1 levels, replacing compaction operations with pointer updates.
Compacting only the last level to the storage. ListDB~\cite{listdb} builds skiplists using the WAL, avoiding costly memtable
    {se,dese}rialization operations. Also, similarly to MioDB, uses pointer operations to update skiplists, avoiding compactions using the
Zipper compaction. Problematically, all these memory-based approaches keep more data in (DRAM or NVM) memory, rendering them much less
cost-effective compared to memory-frugal solutions like RocksDB and vLSM.

\beginbsec{Compaction Strategies}
Other work in LSM KV stores has targeted relevant aspects of compaction strategies. Spooky~\cite{spooky} proposes a hybrid
compaction strategy to reduce I/O amplification without increasing space amplification on the device.  It uses full compaction for the
higher levels of the LSM-tree and incremental compaction for the lower levels.  Although this is not the goal of Spooky, this approach will
increase tail latency because of the full compaction steps at the higher levels of the store. Dostoevsky~\cite{dostoevsky} introduces lazy
leveling that combines tiering and leveling in the same LSM-tree to reduce I/O amplification significantly while sacrificing up to 10\% of
the overall space. Like Spooky, Dostoevsky will result in high tail latency when the N-1 levels become full since full compactions will
occur for each level until the last level. Also, due to tiering compaction, Dostoevsky cannot use incremental compaction for the tiered
levels, increasing exponentially the compaction width.

\beginbsec{Compaction scheduling}
bLSM~\cite{blsm} proposes a scheduling approach to reduce the tail latency of LSM KV stores. It uses a spring and gear algorithm to free a
portion of each level during compaction. bLSM uses full (instead of incremental) compaction, however, stops each compaction step at
scheduled points, to ensure progress at all levels. Silk~\cite{silk} identifies two root causes of write stalls in LSM KV stores: (a) the
$L_0$ to $L_1$ compaction and (b) the scheduling of I/O operations. It proposes a scheduling approach that prioritizes the compaction of the
$L_0$ level to reduce tail latency and defers the compaction of the higher levels to execute when the system is not under heavy load.

ADOC~\cite{adoc} identifies that SILK increases memory consumption up to 22\% and instead tries to reduce tail latency using the overflow
capabilities of incremental compaction by allowing levels to exceed their size limits. As a result, ADOC reduces tail latency by sacrificing
I/O amplification. To prevent excessive I/O amplification overheads, ADOC adjusts the number of compaction threads and batch sizes to reduce
the period when the system is overflowing. However, this requires the server to have available CPU and memory resources.
Calcspar~\cite{calcspar} implements a scheduler for systems deployed in cloud environments where IOPS per second are limited.

Calcspar differentiates between high-priority requests (user-facing requests) and low-priority requests (compaction and prefetching
requests). It prioritizes user-facing requests and opportunistically compacts $L_1$ to $L_2$ levels, while below $L_2$, it defers compaction
for a short period. However, like ADOC, Calcspar might observe high I/O amplification overheads if the system defers compaction for an
extended period. \name{} is compatible with all of the above scheduling approaches and could further reduce tail latency by adopting one of
them based on the deployed environment.

\section{Conclusion}
\label{sec:conclusions}

In this paper, first we analyze the factors that affect tail latency in modern KV stores and the trade-off with I/O amplification
and the amount of memory. We use the notion of compaction chains to understand how different parameters manage this trade-off and we
show that modern designs and existing techniques are not able to optimize at the same time all three aspects: tail latency, I/O
amplification, and memory. Then we present \name{} an LSM KV store that reduces tail latency by reducing both the width and length
of compaction chains, without increasing I/O amplification and the amount of memory required.  \name{} reduces (a) chain width by
using small SSTs and eliminating the tiering compaction and (b) chain length by using a larger growth factor across the first
device device levels ($L_1$, $L_2$) and introducing overlap-aware SSTs (vSSTs).  Compared to RocksDB, \name{} reduces tail latency
by 4.8$\times$ (12.5$\times$) for writes (reads).


\bibliographystyle{plain}
\bibliography{paper}
\end{document}